\documentclass[10pt,conference]{IEEEtran}
\IEEEoverridecommandlockouts
\usepackage{cite}

\usepackage[utf8]{inputenc}
\usepackage[english]{babel}
\usepackage{fancyhdr}
\usepackage{lastpage}

\usepackage{amsmath,amssymb,amsfonts}
\usepackage{algorithmic}
\usepackage{graphicx, subcaption}
\usepackage{textcomp}
\usepackage{xcolor}
\def\BibTeX{{\rm B\kern-.05em{\sc i\kern-.025em b}\kern-.08em
    T\kern-.1667em\lower.7ex\hbox{E}\kern-.125emX}}

\begin{document}

\title{An Ethereum-based Product Identification System for Anti-counterfeits \\
}

\author{\IEEEauthorblockN{Shashank Gupta}
\IEEEauthorblockA{\textit{Department of Computer Science} \\
\textit{University of Kentucky}\\
Kentucky, USA\\
shashank.gupta@uky.edu}

}
\maketitle


\section{Abstract}
Fake products are items that are marketed and sold as genuine, high-quality products but are actually counterfeit or low-quality knock-offs. These products are often designed to closely mimic the appearance and branding of the genuine product in order to deceive consumers into thinking they are purchasing the real thing. Fake products can range from clothing and accessories to electronics and other goods and can be found in a variety of settings, including online marketplaces and brick-and-mortar stores. Blockchain technology can be used to help detect fake products in a few different ways. One of the most common is through the use of smart contracts, which are self-executing contracts with the terms of the agreement between buyer and seller being directly written into lines of code. This allows for a high level of transparency and traceability in supply chain transactions, making it easier to identify and prevent the sale of fake products and the use of unique product identifiers, such as serial numbers or QR codes, that are recorded on the blockchain. This allows consumers to easily verify the authenticity of a product by scanning the code and checking it against the information recorded on the blockchain. In this study, we will use smart contracts to detect fake products and will evaluate based on Gas cost and ethers used for each implementation.  

\section{Introduction}
Fake products, also known as counterfeit or knock-off goods, are items that are marketed and sold as genuine, high-quality products but are actually low-quality imitations. These products can range from clothing and accessories to electronics and other goods and can be found in a variety of settings, including online marketplaces and brick-and-mortar stores. Examples of fake products include designer knock-off handbags, fake designer clothing, and counterfeit electronics with the brand name of a well-known company. These fake products are often designed to closely mimic the appearance and branding of the genuine product in order to deceive consumers into thinking they are purchasing the real thing.

In this project, we propose to design an anti-counterfeit system based on blockchain technology. Blockchain offers several key benefits for an anti-counterfeit system, including its security, decentralization, and transparency properties. By using blockchain, we can create a secure and transparent record of product authentication that is resistant to tampering and fraud. The decentralized nature of blockchain also allows for a more open and transparent system and enables multiple parties to verify the authenticity of a product. These features make blockchain an ideal technology for designing an effective anti-counterfeit system.
 
\section{Motivation}

Counterfeit goods are imitations or knock-offs of products that are created and sold with the intent to deceive consumers into believing they are genuine. These goods can range from high-end luxury items such as designer handbags and watches to everyday items like medications and consumer electronics. The market for counterfeit goods is vast and lucrative, with estimates suggesting that it is worth hundreds of billions of dollars globally. The production and sale of these goods often occur in underground or illicit markets, making it difficult to track and regulate.

The counterfeiting industry is the biggest criminal enterprise in the world, with sales of fake and pirated goods estimated to be worth between \$1.7 trillion and \$4.5 trillion annually. This is higher than the amount earned through either drug or human trafficking. China is responsible for producing around 80\% of these goods, with 60-80\% of them being bought by Americans. These figures demonstrate the significant effects that such illegal trade has on the US economy, business interests, and innovations. \cite{cui2019blockchain}

Counterfeit goods can be dangerous to consumers for a number of reasons. In the case of medications, for example, counterfeit drugs may not contain the active ingredients they claim to, or they may contain harmful additives that can cause serious health problems. Similarly, counterfeit electronics may not meet safety standards and can pose a risk of fire or electrical shock. In addition to the potential risks to consumers, the production and sale of counterfeit goods also harm legitimate businesses and industries. When consumers purchase counterfeit goods, they are essentially giving money to criminals and supporting illegal activities. This can lead to job losses and economic harm to legitimate companies, as well as damage to a brand's reputation.

In addition to the potential risks to consumers, the production and sale of counterfeit goods also harm legitimate businesses and industries. When consumers purchase counterfeit goods, they are essentially giving money to criminals and supporting illegal activities. This can lead to job losses and economic harm to legitimate companies, as well as damage to a brand's reputation. Consumers can protect themselves by purchasing goods from reputable sources and being cautious of deals that seem too good to be true. It is important to know that purchasing counterfeit goods not only puts oneself at risk but also supports illegal activities and harms legitimate businesses.

Overall, preventing counterfeit products is crucial for protecting consumers, supporting legitimate businesses, and safeguarding society as a whole. Governments and law enforcement agencies around the world are working to combat the market for counterfeit goods, but it is also important for individuals to be aware of the issue and avoid purchasing counterfeit products.

\section{Related Work}
One of the potential uses of blockchain technology is supply chain tracking.\cite{8751415, sunny2020supply, cui2019blockchain} By creating an immutable record of transactions using blockchain, companies can track the movement of goods through their supply chain in a transparent and secure way. This can help improve the efficiency of supply chain operations and reduce the risk of fraud. Additionally, the use of blockchain in supply chain tracking can provide greater visibility and accountability for all parties involved in the supply chain, from manufacturers to distributors to retailers. This can help to improve the overall trust and transparency of the supply chain. Overall, the use of blockchain in supply chain tracking has the potential to revolutionize the way that goods are tracked and traced throughout the supply chain.

Another potential use of blockchain technology is identity management.\cite{liu2020blockchain, jacobovitz2016blockchain} By using blockchain to create a digital identity system, individuals would be able to have greater control over their own personal information and would be able to share it with others on their own terms. This could be especially useful in situations where individuals need to prove their identity, such as when applying for a loan or opening a bank account. Additionally, a blockchain-based digital identity system would be more secure than traditional systems, as it would be difficult for an attacker to alter or forge the records on the blockchain. Overall, the use of blockchain in identity management has the potential to improve security and privacy while also making it easier for individuals to prove their identity.

A voting system is another application of blockchain technology. \cite{hjalmarsson2018blockchain} By using blockchain to create a secure, transparent, and auditable voting system, it would be possible to ensure that elections are fair and free from tampering. Additionally, a blockchain-based voting system would allow for real-time tracking and counting of votes, which could help to increase the efficiency and speed of the election process. Furthermore, a blockchain-based voting system could improve voter turnout by making it easier for individuals to cast their ballots, even if they are unable to physically attend a polling place. Overall, the use of blockchain in voting systems has the potential to improve the security, transparency, and accessibility of the election process.

\section{Blockchain}
Blockchain technology is a decentralized, digital ledger that records transactions on multiple computers in a way that makes them tamper-resistant. It is the underlying technology behind cryptocurrencies such as Bitcoin, but it has many other potential uses. Blockchain technology works by allowing multiple parties to contribute data to the ledger, but once that data is added, it cannot be altered or deleted. This creates a permanent and tamper-proof record of transactions, which can be accessed and verified by anyone on the network. This allows for a high level of transparency and trust among the parties involved and makes it difficult for fraudulent activity to go undetected.

One of the key benefits of blockchain technology is its ability to facilitate peer-to-peer transactions without the need for a central authority or intermediary. This allows for more efficient and cost-effective transactions and can help to reduce the risk of fraud and other types of crime.

In a blockchain, data is stored in blocks that are linked together in a chain. Each block contains a number of transactions, and once those transactions are added to the block, they cannot be altered or deleted. This creates a permanent and tamper-proof record of the transactions.
Each block also contains a unique code, known as a "hash," which distinguishes it from other blocks in the chain. This hash is generated using complex algorithms and is based on the data contained in the block, as well as the hash of the previous block in the chain. This creates a secure and unbroken chain of blocks that can be easily verified and traced back to their origin. The data on the blockchain is distributed across the network, with each node on the network having a copy of the entire blockchain. This allows for decentralized storage of the data and ensures that the information on the blockchain is accessible to anyone on the network.

\section{Ethereum}
Ethereum is a decentralized, open-source blockchain platform that allows for the creation of smart contracts and decentralized applications (dapps). It was launched in 2015 by Vitalik Buterin, a programmer and researcher who was involved in the development of the original blockchain technology, upon which Bitcoin is based. It uses a proof-of-work consensus mechanism. This means that in order to add a new block to the blockchain, users must solve complex mathematical expressions, a process known as mining. By doing this, they "prove" that they have put in the computational work to add the block, and this process confirms that the block has been successfully added to the blockchain. As a reward for successfully adding a block, miners are rewarded with ETH, the native cryptocurrency of the Ethereum blockchain.

\section{Goals}
Our goals for this project are as follows:

\textbf{1}. The goal of the project is to create a secure and transparent system for tracking and verifying the authenticity of products using blockchain technology.

\textbf{2}. To make the system user-friendly and accessible, so that consumers can easily verify the authenticity of a product and businesses can participate in the system without needing to have a deep understanding of blockchain technology.

\textbf{3}. To secure product details using a QR code, which will be scanned by consumers to verify the authenticity of a product.

\textbf{4}. To provide security to clients by offering data and transparency about the products they are purchasing.

\textbf{5}. To improve the ability to detect fake products in the marketplace, and to enhance the overall performance of the anti-counterfeit system.

By achieving these goals, we aim to create an effective and secure anti-counterfeit system that will protect consumers, support legitimate businesses, and help to disrupt the market for counterfeit goods. 

\section{Technical Soundness of Approach}
\subsection{Smart Contract}
A smart contract is a self-executing contract with the terms of the agreement between buyer and seller being directly written into lines of code. This code is deployed to the blockchain, where it can be executed automatically when certain predetermined conditions are met.

Smart contracts provide a number of benefits over traditional contracts. Because they are written in code and executed automatically on the blockchain, they can be enforced without the need for a third-party intermediary. This makes the execution of the contract more efficient and cost-effective. Additionally, because the terms of the contract are written into the code, they are transparent and easy to verify, reducing the risk of disputes or misunderstandings.

\subsection{Testnet}
Goerli is a testnet for the Ethereum blockchain. Testnets are test versions of blockchain networks that are used by developers to test and experiment with new features and applications before they are deployed to the main network.

Goerli is a particularly useful testnet for Ethereum developers because it is a "proof-of-authority" network, which means that it is more stable and predictable than other testnets. This makes it a good environment for testing and debugging smart contracts and other Ethereum-based applications.

\subsection{Remix IDE}
One way to write and deploy a smart contract on a testnet like Goerli is to use the Remix IDE. The remix is a web-based Integrated Development Environment (IDE) for the Ethereum platform. It is a tool that allows developers to write, test, and deploy smart contracts on the Ethereum blockchain and then deploy it to a testnet like Goerli for further testing and use.

\subsection{Wallet}
Metamask is a popular browser extension that allows users to interact with the Ethereum blockchain. It serves as a digital wallet for users to store, manage, and transfer their Ether and other Ethereum-based assets. It also allows users to access decentralized applications (dApps) built on the Ethereum platform. It also integrates seamlessly with the web browsers that it is available for, allowing users to easily access their wallets and interact with dApps while browsing the internet.

\section{Problem formulation}
The proposed product tracking system using blockchain technology aims to improve supply chain transparency and traceability. The system assumes that all producers, distributors, and retailers are trusted nodes, and it maintains the status of each product, including the manufacturer of the product, the current owner of the product, and the history of the owners. Each product is assigned a unique product ID, which is used to generate a QR code. This product ID is maintained on the blockchain as the key to a product, allowing for the tracking of its history.

At each status change of a product, the manufacturer, distributor, or retailer will update the new information to the blockchain. When the product is created, it is labeled as "available" on the blockchain. When the product is sold to the distributor or retailer, it remains labeled as "available," and only when it is sold to the end customer is it labeled as "unavailable." Only products labeled as "available" can be sold.

Customers can check the availability and track the history of a product before purchase by scanning the QR code with their mobile device. This allows them to verify the authenticity and provenance of the product, as well as ensure that it has not already been sold to someone else.

Overall, the use of blockchain technology in this product tracking system has the potential to improve supply chain efficiency and trust, as well as give customers greater confidence in the products they purchase. Figure \ref{Fig:flow} and Figure \ref{Fig:diagram}  are shown for displaying the flow and how it improves security.

\section{Methodology}
\subsection{Dummy Ethers}

We signed up to MetaMask to create the wallet. Apart from the mainnet, we manually need to add the Goerli testnet into our wallet. Goerli testnet has a global chain ID 5. Then we signed up on Alchemy and linked our wallet Address to it. Then we used Goerli Faucet to request Dummy Ethers in our connected wallet. It provides 0.5 Ethers on the testnet wallet once in every 24 hours period on request.  

\subsection{Writing Smart Contract}
To use Remix to write a smart contract and deploy it to Goerli, first open the Remix IDE in a web browser. Next, create a new file and write the smart contract code using the Solidity programming language. The next step is to connect your MetaMask wallet to remix IDE. Once the code is written and complied successfully, select "Injected Provider- Metamask" in the environment section. Then choose the account from which you requested dummy ethers and use the deploy button to deploy the smart contract to the testnet.

After the smart contract is deployed to Goerli, it can be accessed and interacted with using a web3 wallet or another Ethereum-compatible tool. This allows developers to test the functionality of the smart contract and ensure that it is working as intended.

\subsection{Problem formulation}
The proposed product tracking system using blockchain technology aims to improve supply chain transparency and traceability. The system assumes that all producers, distributors, and retailers are trusted nodes, and it maintains the status of each product, including the manufacturer of the product, the current owner of the product, and the history of the owners. Each product is assigned a unique product ID, which is used to generate a QR code. This product ID is maintained on the blockchain as the key to a product, allowing for the tracking of its history.

At each status change of a product, the manufacturer, distributor, or retailer will update the new information to the blockchain. When the product is created by the manufacturer, it is labeled as "available" on the blockchain. When the product is sold to the distributor or retailer, it remains labeled as "available," and only when it is sold to the end customer is it labeled as "unavailable." Only products labeled as "available" can be sold.

Customers can check the availability and track the history of a product before purchase by scanning the QR code with their mobile device. This allows them to verify the authenticity and provenance of the product, as well as ensure that it has not already been sold to someone else.

\begin{figure}[ht]
      \includegraphics[width=8cm, height=7cm]{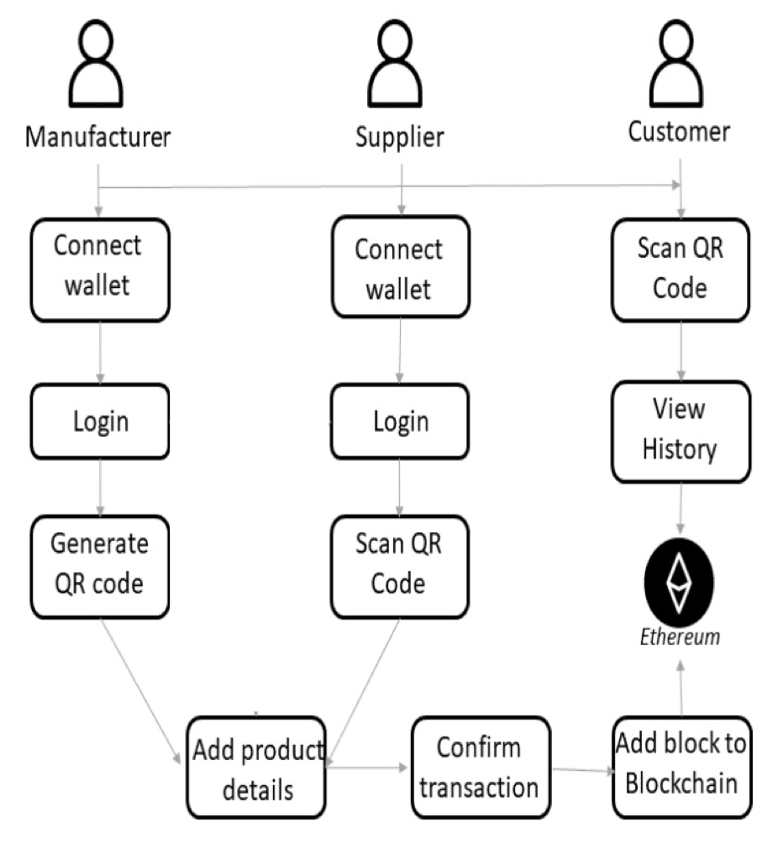}
      \caption{Supply Chain Flow \cite{wasnik2022detection}}
\label{Fig:flow}

\end{figure}

\begin{figure}[ht]
      \includegraphics[width=8cm, height=7cm]{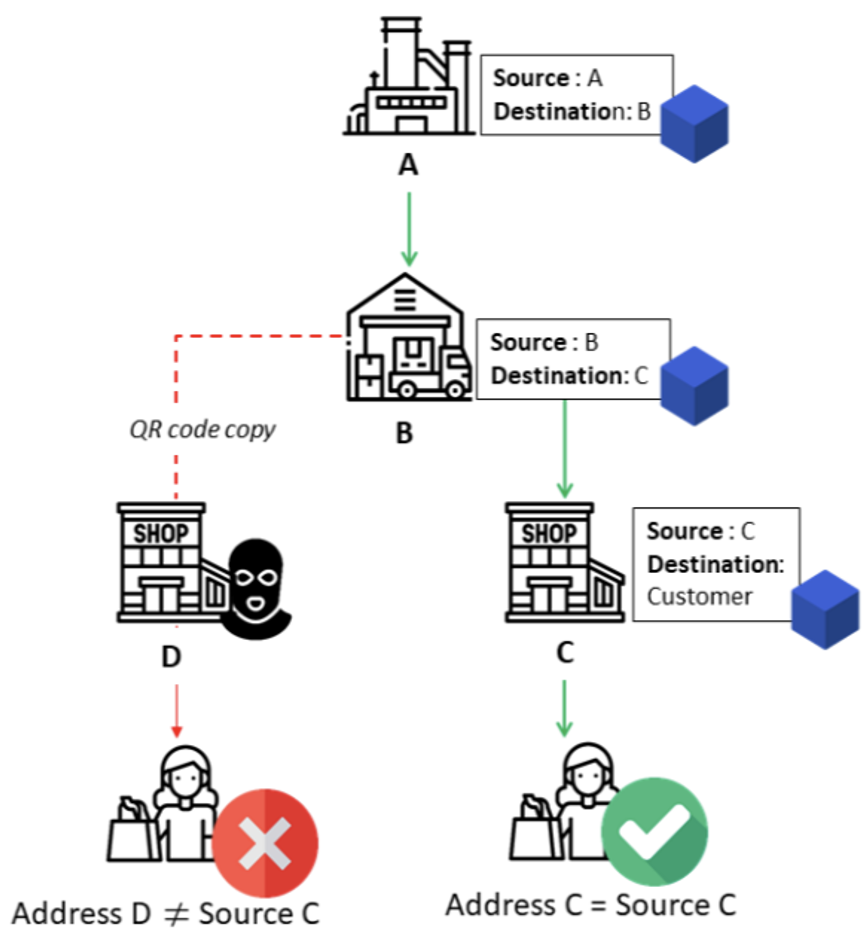}
      \caption{Detection of Fake Product \cite{wasnik2022detection}}
\label{Fig:diagram}

\end{figure}

Overall, the use of blockchain technology in this product tracking system has the potential to improve supply chain efficiency and trust, as well as give customers greater confidence in the products they purchase. Figure \ref{Fig:flow} and \ref{Fig:diagram} show the flow and how it improves security.



\section{Evaluation}
\begin{figure}[ht]
      \includegraphics[width=8cm, height=10cm]{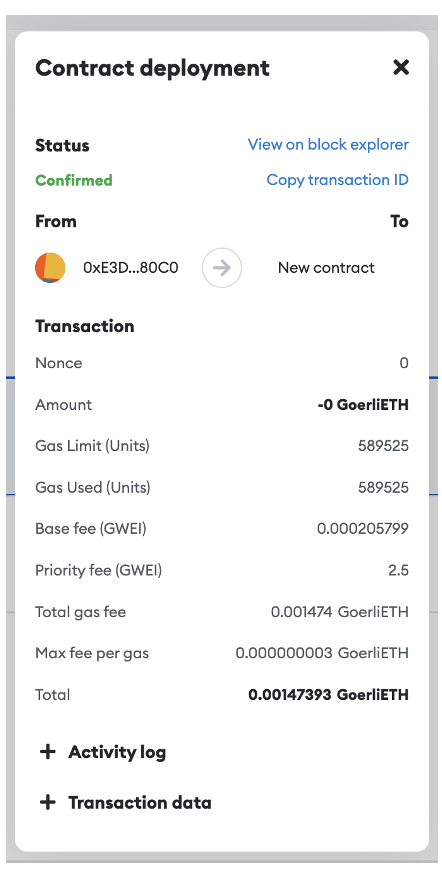}
      \caption{Smart contract deployment cost}
\label{Fig:contract-cost}
\end{figure}

\begin{figure}[ht]
      \includegraphics[width=8cm, height=10cm]{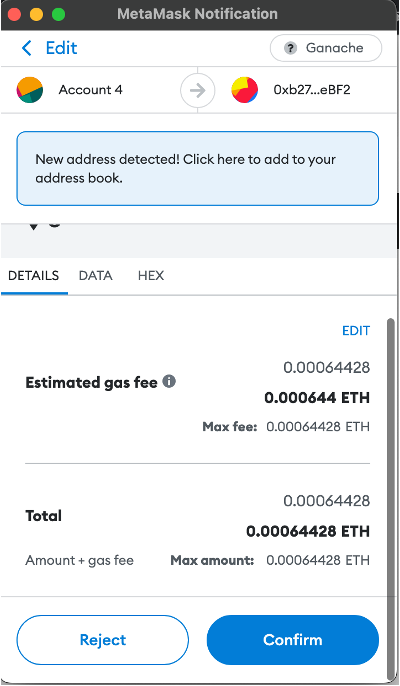}
      \caption{Product Registration Cost}
\label{Fig: Product Registration Cost}
\end{figure}

\begin{figure}[ht]
      \includegraphics[width=8cm, height=10cm]{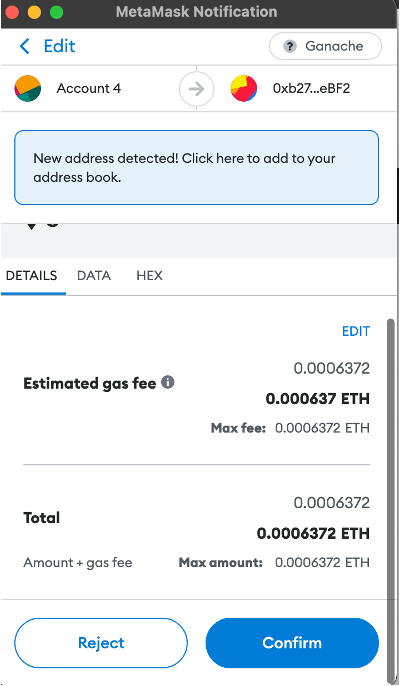}
      \caption{Product Registration Cost}
\label{Fig: Product Selling Cost}
\end{figure}

In order to determine the effectiveness of our implementation, we evaluate it using several metrics related to the cost of running transactions on the Ethereum network. One of these metrics is the gas price, which is the amount of Ether that must be paid to a miner for every unit of gas consumed by a transaction. This is an important factor because it directly affects the cost of running a transaction on the Ethereum network. Another metric that we use to evaluate our implementation is the gas cost per transaction, which is the total amount of gas consumed by a transaction divided by the number of transactions performed. This metric allows us to compare the efficiency of different implementations and identify ways to reduce the cost of running transactions on the Ethereum network. Finally, we also consider the ETH cost, which is the total amount of Ether spent on transactions. This metric provides a broader view of the overall cost of using the Ethereum network and helps us evaluate the overall efficiency of our implementation. See Figure \ref{Fig:contract-cost}, \ref{Fig: Product Registration Cost}, and \ref{Fig: Product Selling Cost} for our evaluation results.

\section{Conclusion}
As we can see from the evaluation cost, deployment of a Smart contract is costly as compared to product registration and product selling. But Smart contract deployment is only a one-time fee whereas product selling and registration is a continuous process. The total cost of the system is 0.00064428 ETH

\bibliographystyle{ACM-Reference-Format}
\bibliography{main}


\begin{thebibliography}{7}


\ifx \showCODEN    \undefined \def \showCODEN     #1{\unskip}     \fi
\ifx \showDOI      \undefined \def \showDOI       #1{#1}\fi
\ifx \showISBNx    \undefined \def \showISBNx     #1{\unskip}     \fi
\ifx \showISBNxiii \undefined \def \showISBNxiii  #1{\unskip}     \fi
\ifx \showISSN     \undefined \def \showISSN      #1{\unskip}     \fi
\ifx \showLCCN     \undefined \def \showLCCN      #1{\unskip}     \fi
\ifx \shownote     \undefined \def \shownote      #1{#1}          \fi
\ifx \showarticletitle \undefined \def \showarticletitle #1{#1}   \fi
\ifx \showURL      \undefined \def \showURL       {\relax}        \fi
\providecommand\bibfield[2]{#2}
\providecommand\bibinfo[2]{#2}
\providecommand\natexlab[1]{#1}
\providecommand\showeprint[2][]{arXiv:#2}

\bibitem[Cui et~al\mbox{.}(2019)]%
        {cui2019blockchain}
\bibfield{author}{\bibinfo{person}{Pinchen Cui}, \bibinfo{person}{Julie Dixon},
  \bibinfo{person}{Ujjwal Guin}, {and} \bibinfo{person}{Daniel Dimase}.}
  \bibinfo{year}{2019}\natexlab{}.
\newblock \showarticletitle{A blockchain-based framework for supply chain
  provenance}.
\newblock \bibinfo{journal}{\emph{IEEE Access}}  \bibinfo{volume}{7}
  (\bibinfo{year}{2019}), \bibinfo{pages}{157113--157125}.
\newblock


\bibitem[Hj{\'a}lmarsson et~al\mbox{.}(2018)]%
        {hjalmarsson2018blockchain}
\bibfield{author}{\bibinfo{person}{Fri~rik Hj{\'a}lmarsson},
  \bibinfo{person}{Gunnlaugur~K Hrei~arsson}, \bibinfo{person}{Mohammad
  Hamdaqa}, {and} \bibinfo{person}{G{\'\i}sli Hj{\'a}lmt{\`y}sson}.}
  \bibinfo{year}{2018}\natexlab{}.
\newblock \showarticletitle{Blockchain-based e-voting system}. In
  \bibinfo{booktitle}{\emph{2018 IEEE 11th international conference on cloud
  computing (CLOUD)}}. IEEE, \bibinfo{pages}{983--986}.
\newblock


\bibitem[Jacobovitz(2016)]%
        {jacobovitz2016blockchain}
\bibfield{author}{\bibinfo{person}{Ori Jacobovitz}.}
  \bibinfo{year}{2016}\natexlab{}.
\newblock \showarticletitle{Blockchain for identity management}.
\newblock \bibinfo{journal}{\emph{The Lynne and William Frankel Center for
  Computer Science Department of Computer Science. Ben-Gurion University, Beer
  Sheva}}  \bibinfo{volume}{1} (\bibinfo{year}{2016}), \bibinfo{pages}{9}.
\newblock


\bibitem[Liu et~al\mbox{.}(2020)]%
        {liu2020blockchain}
\bibfield{author}{\bibinfo{person}{Yang Liu}, \bibinfo{person}{Debiao He},
  \bibinfo{person}{Mohammad~S Obaidat}, \bibinfo{person}{Neeraj Kumar},
  \bibinfo{person}{Muhammad~Khurram Khan}, {and}
  \bibinfo{person}{Kim-Kwang~Raymond Choo}.} \bibinfo{year}{2020}\natexlab{}.
\newblock \showarticletitle{Blockchain-based identity management systems: A
  review}.
\newblock \bibinfo{journal}{\emph{Journal of network and computer
  applications}}  \bibinfo{volume}{166} (\bibinfo{year}{2020}),
  \bibinfo{pages}{102731}.
\newblock


\bibitem[Niya et~al\mbox{.}(2019)]%
        {8751415}
\bibfield{author}{\bibinfo{person}{Sina~Rafati Niya}, \bibinfo{person}{Danijel
  Dordevic}, \bibinfo{person}{Atif~Ghulam Nabi}, \bibinfo{person}{Tanbir Mann},
  {and} \bibinfo{person}{Burkhard Stiller}.} \bibinfo{year}{2019}\natexlab{}.
\newblock \showarticletitle{A Platform-independent, Generic-purpose, and
  Blockchain-based Supply Chain Tracking}. In \bibinfo{booktitle}{\emph{2019
  IEEE International Conference on Blockchain and Cryptocurrency (ICBC)}}.
  \bibinfo{pages}{11--12}.
\newblock
\urldef\tempurl%
\url{https://doi.org/10.1109/BLOC.2019.8751415}
\showDOI{\tempurl}


\bibitem[Sunny et~al\mbox{.}(2020)]%
        {sunny2020supply}
\bibfield{author}{\bibinfo{person}{Justin Sunny}, \bibinfo{person}{Naveen
  Undralla}, {and} \bibinfo{person}{V~Madhusudanan Pillai}.}
  \bibinfo{year}{2020}\natexlab{}.
\newblock \showarticletitle{Supply chain transparency through blockchain-based
  traceability: An overview with demonstration}.
\newblock \bibinfo{journal}{\emph{Computers \& Industrial Engineering}}
  \bibinfo{volume}{150} (\bibinfo{year}{2020}), \bibinfo{pages}{106895}.
\newblock


\bibitem[Wasnik et~al\mbox{.}(2022)]%
        {wasnik2022detection}
\bibfield{author}{\bibinfo{person}{Kunal Wasnik}, \bibinfo{person}{Isha
  Sondawle}, \bibinfo{person}{Rushikesh Wani}, {and} \bibinfo{person}{Namita
  Pulgam}.} \bibinfo{year}{2022}\natexlab{}.
\newblock \showarticletitle{Detection of Counterfeit Products using
  Blockchain}. In \bibinfo{booktitle}{\emph{ITM Web of Conferences}},
  Vol.~\bibinfo{volume}{44}. EDP Sciences, \bibinfo{pages}{03015}.
\newblock


\end{thebibliography}
\end{document}